\documentclass[%
 aip,
 reprint,%
]{revtex4-1}

\usepackage{graphicx}
\usepackage{dcolumn}
\usepackage{bm}

\usepackage[outdir=./]{epstopdf}
\usepackage{hyperref}
\usepackage{ascmac}
\usepackage{amsmath,amssymb}
\usepackage{indentfirst}
\usepackage{enumitem}
\usepackage{comment}
\usepackage{color}
\usepackage{subcaption}
\usepackage{footnpag}

\usepackage[utf8]{inputenc}
\usepackage[T1]{fontenc}
\usepackage{mathptmx}
\usepackage{etoolbox}
\usepackage{physics}
\usepackage[dvipsnames]{xcolor}
\usepackage{svg}
\usepackage{import}

\makeatletter
\def\@email#1#2{%
 \endgroup
 
 \patchcmd{\titleblock@produce}
  {\frontmatter@RRAPformat}
  {\frontmatter@RRAPformat{\produce@RRAP{*#1\href{mailto:#2}{#2}}}\frontmatter@RRAPformat}
  {}{}
}%
\makeatother

\begin{document}

\renewcommand{\thefootnote}{\arabic{footnote}}

\preprint{AIP/123-QED}

\title{
Applicability criteria of proper charge neutrality and special relativistic MHD models extended by two-fluid effects
}
\author{Shuntaro Yoshino}
\email{shuntaro.yoshino.r2@dc.tohoku.ac.jp}
 \affiliation{Graduate School of Information Science, Tohoku University, Sedai 980-8579, Japan}
\author{Makoto Hirota}%

\author{Yuji Hattori}
\affiliation{%
Institute of Fluid Sciences, Tohoku University, Sedai 980-8577, Japan
}%

\date{\today}

\begin{abstract}
The applicability of relativistic magnetohydrodynamics (RMHD) and its generalization to two-fluid models (including the Hall and inertial effects) is systematically investigated by using the method of dominant balance in the two-fluid equations. Although proper charge neutrality or quasi-neutrality is the key assumption for all MHD models, this condition is difficult to be met when both relativistic and inertial effects are taken into account. The range of application for each MHD model is illustrated in the space of dimensionless scale parameters. Moreover, the number of field variables of relativistic Hall MHD (RHMHD) is shown to be greater than that of RMHD and Hall MHD. Nevertheless, the RHMHD equations may be solved at a lower computational cost than RMHD in the limit of cold plasma, since root-finding algorithm, which is the most time-consuming part of the RMHD code, is no longer required to compute the primitive variables.
\end{abstract}

\maketitle

\section{Introduction}
Magnetized plasmas subject to relativistic effects are common in various high-energy celestial bodies, such as pulsar, black hole magnetosphere, corona of accretion disk, jet from active galactic nucleus and gamma-ray burst. Relativistic magnetohydrodynamics (RMHD) has been used in theoretical and numerical studies as a model to analyze the macroscopic motion of relativistic magnetized plasmas. Since RMHD ignores the microscopic scales of such as inertial length and gyro-radius, it is well known that the magnetic field is frozen in the plasma motion in the zero resistivity limit. Therefore, RMHD is inappropriate for dealing with magnetic reconnection~\cite{uzdensky_magnetic_2011, hoshino_relativistic_2012} at least in the microscopic region where magnetic field lines reconnect. 
In many cases, magnetic reconnection is a key process in which magnetic energy is efficiently converted into kinetic and thermal energy. In addition, RMHD becomes invalid 
in the limit of low plasma density or weak magnetic field. The Vlasov-Maxwell equations, on the other hand, are based on first principles and have been solved by Particle-In-Cell (PIC) simulations in recent years. 
However, this direct approach is the most computationally expensive and these kinetic models are difficult to solve analytically. Thus, there is a demand for intermediate models which bridge the gap between RMHD and kinetic ones. In this study, we focus on extended RMHD models that include the two-fluid effects, which are expected to be more widely applicable than RMHD 
while maintaining a moderate computational cost.

In non-relativistic electron-ion plasmas, a model~\cite{spitzer1956, Lust} including the two-fluid effects (i.e., the Hall effect and the electron inertia effect) is called extended MHD (XMHD) in the recent literature~\cite{kimura2014energy}. The XMHD equations are derived from the two-fluid equations by imposing the quasi-neutrality (QN) condition, which approximately eliminates microscopic motions such as plasma oscillation and cyclotron oscillation. XMHD is also shown to have a Hamiltonian structure which conserves canonical vorticities (instead of magnetic flux)~\cite{keramidas_charidakos_action_2014, Abdelhamid_2015, LINGAM2016, hirota_linearized_2021}.
Due to the electron-inertia effect, magnetic reconnection can occur 
even in the zero resistivity limit~\cite{Dungey, SPEISER1970613, ottaviani_nonlinear_1993}. 

Furthermore, the Hall effect is well-known for significantly enhancing the reconnection speed, according to the Global Environment Modeling (GEM) Reconnection Challenge~\cite{birn_geospace_2001}. Since the electron-inertia effect manifests itself on an even smaller scale than the Hall effect, Hall MHD~\cite{Lighthill1960} is often used as well, neglecting only the electron-inertia effect. 
In the case of electron-positron plasmas, the Hall effect vanishes and only the inertial effect remains, so this model is called inertial MHD (IMHD)~\cite{kimura2014energy}.

It is natural to assume that there are also some MHD models that include the relativistic effects alongside the two-fluid effects. Such an extension of RMHD in electron-positron plasma was explored early on the literature~\cite{lichnerowicz_relativistic_1967, anile_relativistic_2005}. 
Additionally, the extension of generalized Ohm's law was attempted and applied to pulsar magnetosphere\cite{ardavan_magnetospheric_1976}\footnote{Generalized relativistic Ohm's law is also proposed by \cite{pegoraro_generalised_2015} in a different way.}. Koide~\cite{Koide2009} derived a generalized RMHD model from the relativistic two-fluid equations by imposing the proper charge neutrality (PCN) condition, which will be referred to as relativistic extended MHD (RXMHD) in this paper. A variational principle of RXMHD was later proposed by Kawazura et al. \cite{Kawazura2017}. The general relativistic version of RXMHD was presented by Koide\cite{koide_generalized_2009} and 
by Comisso and Asenjo \cite{comisso_collisionless_2018} using a covariant form.
RXMHD was applied to relativistic collisionless magnetic reconnection~\cite{comisso2014}. Relativistic Hall MHD (RHMHD) is similarly obtained by neglecting electron-inertia, and its properties have been studied by Kawazura~\cite{kawazura2017Hall,Kawazura2023}. However, for the QN condition to hold in non-relativistic MHD, the flow velocity must be sufficiently slower than the speed of light. Moreover, as will be clarified in this paper, the PCN condition in RMHD actually holds in the limit of neglecting the two-fluid effects. Therefore, hybrid models which include both the relativistic and two-fluid effects may violate the charge neutrality condition, requiring careful consideration of the applicability of RXMHD (and RHMHD).
In fact, all models bearing the name "MHD" assume either QN or PCN a priori. Once these neutrality conditions (i.e., single-fluid approximation) fail, we should solve the two-fluid equations or kinetic models directly.
For example, the two-fluid model for relativistic electron-positron plasmas ~\cite{barkov_multidimensional_2014} has been solved numerically and applied to the simulations of magnetic reconnection~\cite{zenitani_two-fluid_2009}.

In this study, starting from the relativistic two-fluid equations, we systematically reproduce various MHD models (including RXMHD) using the method of dominant balance~\cite{White} and theoretically illustrate their scopes of application. Since there are too many dimensionless parameters in the original two-fluid equations, we will not explore all cases but focus only on the realm of MHD where the MHD balances hold; the MHD terms are not negligible but dominant. Specifically, we consider a situation in which the Lorentz force ($\bm{J}\times\bm{B}$ term) is dominant in the equation of motion for the center-of-mass velocity of 
the two fluids. If the pressure term or the electric force is far more dominant than the Lorentz force, the MHD model is unlikely to be applicable~\cite{toma_electromotive_2014}. Therefore, in order to make nonessential parameters invisible and highlight only the dominant terms, the plasma pressure and the external electric field will be ignored from the beginning. The MHD models are finally classified in terms of three dimensionless parameters corresponding to the scales of the plasma density, the flow velocity, and the external magnetic field. Furthermore, the dimensionless parameters can be reduced to two if the flow velocity is assumed to be on the same order of the Alfv\'en velocity. The applicability of the various MHD models will be visualized in this parameter space, supposing that a dimensionless coefficient before each term is considered negligible if it is less than, say, $10^{-4}$. For these relativistic and two-fluidic MHD models, we will write them in the form of a dynamical system $\partial_tu=F(u)$ and identify the number of the time-evolving field variables $u$. We will show that RHMHD has more variables than HMHD and RMHD. In the case of RMHD, the right-hand side $F(u)$ is notorious for being an implicit function of $u$, which requires extra computational cost~\cite{komissarov_godunov-type_1999}. RHMHD will be shown to resolve this problem of RMHD, although the number of variables increases.

\section{Basic equations \label{sec: basic}}

\subsection{Two-fluid equations}

We denote the Minkowski spacetime of the reference frame by
\begin{align}
x^\mu :=(ct, x, y, z) = (x^0, x^i), \hspace{10pt} x_\mu := (-ct, x, y, z) = (x_0, x_i),   
\end{align}
where $c$ is the speed of light and the Minkowski metric tensor is
\begin{math}
\mathrm{diag}(-1, 1, 1, 1)
\end{math}. 
The partial derivatives will be shortly denoted by
$\partial_{\mu} ={\partial}/{\partial}{x^{\mu}}$ and 
$\partial^{\mu} ={\partial}/{\partial}{x_{\mu}}$.
The proper four-velocity is defined as
\begin{align}
 U^\mu:=
 (\gamma, \gamma{\vb{v}/c}), 
\hspace{10pt} 
U_\mu :=
(-\gamma, \gamma{\vb{v}/c}),
\end{align}
where $\vb{v}$ is the reference-frame three-velocity (called simply "velocity"), 
and $\gamma:=1/\sqrt{1-|{\vb{v}|^2/c^2}}$ is the Lorentz factor.
In this paper, Greek indices ($\mu=0,1,2,3$) denote the time-space (4D) components, while
Roman indices ($v^i=v_i$, $i=1,2,3$) or bold faces ($\vb{v}$) denote the spatial (3D) components.
The Einstein summation convention will be used in what follows.

Momentarily Co-moving Reference Frame (MCRF) refers to the inertial frame co-moving with particles. Physical quantities of relativistic fluid are said to be "proper" when they are observed in the frame co-moving with the velocity $\vb{v}$. Therefore, the proper number density is given by $N=n/\gamma$, when $n$ is the number density in the reference-frame.

In this study, we start with the special-relativistic fluid equations for both positively and negatively charged gases, where dissipation due to collision is neglected for simplicity. Namely, we ignore resistivity and viscosity.
The equations of motion, the continuity equations and Maxwell's equations are written as
\begin{align}
\label{motion}
\partial_\nu (h_\pm N_{\pm}U_{\pm}^{\mu} U_{\pm}^{\nu})
&=-\partial^\mu p_{\pm} \pm
ecN_{\pm}U_{\pm}^{\nu}{F^{\mu}}_{\nu}, \\
\label{num_flux}
\partial_\nu(N_\pm U^\nu_\pm) &= 0,  \\
\partial_\nu {}^\ast F^{\mu \nu} &= 0, \\
\label{Max2}
\partial_\nu F^{\mu \nu} &= \mu_0 J^\mu,  
\end{align}
where the subscripts plus $(+)$ and minus $(-)$ indicate that they are the quantities for positively and negatively charged particles, respectively. Moreover,
$h_\pm$ is the entropy per unit particle,
$p_\pm$ is the pressure, 
$F^{\mu\nu}$ is the electromagnetic field tensor,
${}^\ast F^{\mu \nu}$ is the Hodge dual tensor of $F^{\mu\nu}$,
and
$J^\mu:=ec(N_+{U_+}^\mu- N_{-}{U_-}^\mu)$ is the four-current. 
The governing equations, \eqref{motion} to \eqref{Max2}, are called the two-fluid equations.
We use the SI unit system; $\mu_0$ is the vacuum magnetic permeability,
and $e$ is the elementary charge.

Maxwell's equations are also expressed in $3 + 1$ form as
\begin{eqnarray}
\label{E}
\partial_j {E^j}/c &=& \mu_0 ec \tilde{n}, \\
\label{M}
\partial_j {B^j}/c &=& 0, \\
\label{F}
\epsilon^i {}_{jk} \partial^j E^k /c &=& -\partial_0 B^i, \\
\label{A_M}
\epsilon^i {}_{jk} \partial^j B^k &=& \mu_0 J^i -\partial_0 E^i /c,
\end{eqnarray}
where $E^i$ is the electric field, 
$B^i$ is the magnetic field, 
$e \tilde{n} = e (n_+ - n_-)$ is the charge density, and
$\epsilon_{ijk}$ is the Levi-Civita symbol.

In this paper, the four potential $A^\mu=(\phi/c,A^i)$ is also introduced to express the electromagnetic field and we employ the Coulomb gauge $\partial_jA^j=0$. Maxwell's equations are then transformed into
\begin{align}
    -\partial_j\partial_j\phi=&\mu_0ec^2\tilde{n},\label{Poisson}\\
      \frac{1}{c^2}\partial_t\partial_tA^i+\partial_i\partial_jA^j-\partial_j\partial_jA^i=&-\frac{1}{c^2}\partial_i(\partial_t\phi)+\mu_0J^i.\label{Ampere}
\end{align}

\subsection{Transformation into MHD variables}

Let us rewrite Eq. \eqref{motion} and Eq. \eqref{num_flux} in terms of MHD variables without any approximation. 
For this purpose, we define the mass-weighted average of number density $ n $, the center-of-mass velocity $v^i $, the number density difference $ \tilde{n} $, and the velocity $u^i$ for the electric current as follows.
\begin{eqnarray}
    \label{n}
    n &:=& \frac{m_+ n_+ + m_- n_-}{m_+ + m_-}, \hspace{10pt} v^i :=  \frac{m_+ n_+ v^i_+ + m_- n_- v^i_-}{(m_+ + m_-)n}, \notag \\
    \tilde{n} &:=& n_+ - n_-, \hspace{10pt} u^i := \frac{n_+ v^i_+ - n_- v^i_-}{n},
\end{eqnarray}
where $m_\pm$ is the mass of particle. Moreover, these variables are associated with four-dimensional center-of-mass flux (divided by $c$),
\begin{eqnarray}
\label{Q}
Q^\mu &=&\qty(n, \ n\frac{v^i}{c}):=\frac{m_+ N_+ {U^{\mu}_+} + m_- N_- {U^{\mu}_-}  }{m_++m_-} , 
\end{eqnarray}
and four-dimensional current (with the same dimension as $Q^\mu$),
\begin{eqnarray}
\label{K}
K^\mu &=&\frac{J^\mu}{ec}=\qty(\tilde{n}, \ n \frac{u^i}{c}):=  N_+ {U^{\mu}_+} - N_- {U^{\mu}_-}.
\end{eqnarray}

The MHD equations are sometimes called the single-fluid model, assuming that the two species of charged fluid move together approximately; $\tilde{n}\ll n$ and $|\vb{u}|\ll|\vb{v}|$.

By denoting \eqref{motion} for the positive and negative species by \eqref{motion}$_+$ and \eqref{motion}$_-$ respectively, the equation of center-of-mass motion is obtained from the sum \eqref{motion}$_++$\eqref{motion}$_-$ as follows
\begin{eqnarray}
\label{mass_flux}
&&\hspace{-10pt} mc^2\partial_\nu \qty[ \ f \ Q^\mu Q^\nu 
+\mu^2 \tilde{f} \  (Q^\mu K^\nu + K^\mu Q^\nu \bigr )
 + \mu^2 {f^\prime}  \ K^\mu K^\nu \ ] \notag \\
&& = -\partial^\mu p + ec K^\nu F^\mu{}_\nu.
\end{eqnarray}
On the other hand, generalized Ohm's law is obtained from $[m_-$\eqref{motion}$_+-m_+$\eqref{motion}$]/(m_++m_-)$,
\begin{eqnarray}
 \label{current}
&&\hspace{-10pt}  mc^2\partial_\nu \bigl [ \ \mu^2 \tilde{f}  \ Q^\mu Q^\nu  
 + \mu^2 {f^\prime} \  (Q^\mu K^\nu + K^\mu Q^\nu \bigr ) \notag \\
&& \hspace{62.5pt} + \mu^2 \bigl (-\tilde{\mu}{f^\prime} +\mu^2 \tilde{f} \bigr ) \ K^\mu K^\nu \ \bigr ] \notag \\ 
&&= \frac{\mu}{2} \partial^\mu p +\frac{1}{2} \partial^\mu \tilde{p} + ec\bigl (Q^\nu - \tilde{\mu} K^\nu \bigr) F^\mu {}_\nu,
\end{eqnarray}
where the following abbreviations are used
\begin{eqnarray}
\label{def_m}
m :=  m_+ + m_-,\quad
\mu^2 := \frac{m_+ m_-}{m^2},\quad
\tilde{\mu} :=\frac{ m_+ - m_-}{m},\\
\label{def_p}
p := p_+ + p_-, \hspace{10pt}
\tilde{p} := p_+ - p_-,
\end{eqnarray}
\begin{eqnarray}
f_\pm &:=& \frac{h_\pm}{N_\pm m_\pm c^2}, \\
\label{def_tilde_f}
\tilde{f} &:=& f_+ - f_-, \\ 
\label{def_f}
f &:=& \frac{1}{m} \qty( m_+ f_+ + m_- f_- ), \\
\label{def_fprime}
f^\prime &:=& \frac{1}{m} \qty( m_- f_+ + m_+ f_-) = f - \tilde{\mu}\tilde{f}.
\end{eqnarray}
Both the classical and relativistic MHD equations are derived by neglecting $\tilde{f}$ owing to $\tilde{f}\ll f$ (which then leads to $f^\prime=f$). Therefore, the orders of $f$ and  $\tilde{f}$ are important for the validity of the MHD approximation.

Similarly, we obtain the conservation law of mass density 
\begin{eqnarray}
\label{par_Q}
\partial_\nu Q^\nu = 0
\end{eqnarray}
from $[m_+$\eqref{num_flux}$_+ + m_-$\eqref{num_flux}$_-]/m$, and the conservation law of charge density
\begin{eqnarray}
 \label{par_K}
\partial_\nu K^\nu = 0
\end{eqnarray}
from \eqref{num_flux}$_+ -$\eqref{num_flux}$_-$.

\subsection{Assumption of cold plasma}

The validity of the MHD approximation primarily relies on $\tilde{n}\ll n$, $|\vb{u}|\ll|\vb{v}|$ and $\tilde{f}\ll f$ being sufficiently fulfilled. To focus on this topic, we neglect the pressure terms (i.e., the cold plasma approximation) in what follows because they simply appear as additional terms and make the governing equations lengthy. 
Therefore, $p =\tilde{p}= 0$ and $h_\pm = m_\pm c^2$ are assumed. Then, \eqref{def_tilde_f} and \eqref{def_f} are reduced to 
\begin{eqnarray} 
f=\frac{1}{m}\left(\frac{m_+}{N_+}+\frac{m_-}{N_-}\right), \quad 
 \tilde{f} = \frac{1}{N_+}-\frac{1}{N_-}.
\end{eqnarray}
Using the relations,
\begin{align}
         N_\pm=&\sqrt{-N_\pm N_\pm U_\pm^\alpha U_{\pm\alpha}}\nonumber\\
      =&\sqrt{-\left(Q^\alpha Q_\alpha+\frac{m_\mp^2}{m^2}K^\alpha K_\alpha\right)
      \mp\frac{m_\mp}{m}\left(Q^\alpha K_\alpha+K^\alpha Q_\alpha\right)},
\end{align}
we can express $f$ and $\tilde{f}$ in terms of $Q^\mu$ and $K^\mu$ (in a very complicated way). 
In Maxwell's equations, the electromagnetic field $F^{\mu\nu}$ is generated by $J^\mu$, which is $ecK^\mu$.
Therefore, the two-fluid equations are fully expressed by the MHD variables, $Q^\mu$, $K^\mu$ and $F^{\mu\nu}$. 

Let us clarify the number of field variables in the two-fluid equations. From the definitions given above, the equations \eqref{mass_flux}, \eqref{current}, \eqref{par_Q} and \eqref{par_K} clearly describe the time evolution of the 8 variables $Q^\mu$ and $K^\mu$ (which correspond to $n$, $v^i$, $\tilde{n}$ and $u^i$). Maxwell's equations provide the time evolution of the 6 variables $E^i$ and $B^i$ (which is $F^{\mu\nu}$), but they must be solved under the 
 two constraints \eqref{E} and \eqref{M} (which include no time derivative). 
 In fact, we can eliminate the variable $\tilde{n}$ because $\tilde{n}$ is uniquely determined by $E^i$ via \eqref{E}, and the charge conservation law \eqref{par_K} is automatically satisfied by \eqref{E} and \eqref{A_M}.
 Therefore, in the cold plasma approximation, the two-fluid equations constitute a dynamical system of 13 field variables under 1 constraint in total. In a sense, the {\it degree of freedom} is $13-1=12$. Even when $A^i$ is used instead of $B^i$, the Coulomb gauge $\partial_iA^i=0$ is imposed instead of $\partial_iB^i=0$ and the {\it degree of freedom} is the same.
 Reducing the number of field variables is one of the major purposes of the following MHD approximation.

\section{Dominant balance \label{sec: scale}}

To derive reduced models from the two-fluid equations systematically, we first normalize all terms in the equations and consider the dominant balances that are suitable for magnetized plasma.

\subsection{Normalization}

We normalize all the equations by introducing 8 representative scales (with subscript $\star$) as follows
\begin{eqnarray}
\hat{n}=\frac{n}{n_\star},\quad
\hat{v}^i=\frac{v^i}{v_\star},\quad
\hat{\tilde{n}}=\frac{\tilde{n}}{\tilde{n}_\star},\quad
\hat{u}^i=\frac{u^i}{u_\star},\notag\\
\hat{B}^i=\frac{B^i}{B_\star},\quad
\hat{\phi}^i=\frac{\phi}{\phi_\star},\quad
\hat{x}^i=\frac{x^i}{L_\star},\quad
\hat{t}=\frac{t}{T_\star},
\end{eqnarray}
where we have introduced the common scale for all three-dimensional components of vector fields (i.e., $v^1\sim v^2\sim v^3$) for simplicity. Note that $L_\star$ and $T_\star$ are the representative spatial and temporal scales, respectively, of plasma dynamics that we are interested in.

The conservation law of mass \eqref{par_Q} is written in terms of the normalized quantities (with the hat symbol) as
\begin{eqnarray}
\frac{L_\star}{v_\star T_\star} \hat{\partial_0} \hat{Q^0} + \hat{\partial_i} \hat{Q^i} &=& 0.
\end{eqnarray}
Except when we consider the special cases (such as steady solution or incompressible limit), the two terms on the left hand side balance each other. First of all, we assume this balance as usual,
\begin{eqnarray}
\mbox{Balance 1: }L_\star=v_\star T_\star.\label{balance1}
\end{eqnarray}
Since this balance is merely a relation among scale parameters, it should be actually interpreted as $L_\star\sim v_\star T_\star$ or $O(L_\star)=O(v_\star T_\star)$. However, in this paper, the equality ``$=$'' will be used to reduce the number of the scale parameters by imposing this balance.

Next, we consider the Poisson equation \eqref{Poisson} which is normalized to
\begin{align}
-\frac{\phi_\star}{L_\star^2}\hat{\partial}_j\hat{\partial}_j\hat{\phi}=\mu_0ec^2\tilde{n}_\star\hat{\tilde{n}}.
\end{align}
and assume that there is no externally-applied electrostatic potential (e.g., $\phi\rightarrow0$ at infinity). Then, $\hat{\phi}$ is generated only by the charge density of plasma itself via this equation, and it is natural to assume the balance between the left and right hand sides,
\begin{eqnarray}
\mbox{Balance 2: }\phi_\star=\mu_0 ec^2 \tilde{n}_\star L_\star^2. \label{balance2}
\end{eqnarray}
Since the two balances 1 and 2 are assumed among the eight representative scales, let us define 5 dimensionless parameters for later use as follows
\begin{align}
\label{inertia}
 \epsilon :=& \frac{1}{L_\star} \sqrt{\frac{m}{\mu_0 n_\star e^2}},\\
\label{sigma}
\sigma :=&  {B_\star^2}/{(\mu_0 m n_\star c^2)}, \\
\label{beta}
\beta_\star :=& v_\star/c, \\
\alpha :=& \tilde{n}_\star/n_\star, \\
\epsilon_m :=& u_\star/v_\star,
\end{align}
where $\epsilon$ denotes the normalized inertial length and $\sigma$
is called the magnetization parameter. As we have mentioned earlier, the smallness of $\alpha$ and $\epsilon_m$ will be essential for the MHD approximation.

Using $\hat{A}^i=A^i/(B_\star L_\star)$, Ampere-Maxwell's law \eqref{Ampere} is normalized as
\begin{align}
      \beta_\star^2\hat{\partial}_0\hat{\partial}_0\hat{A}^i
      +\hat{\partial}_i\hat{\partial}_j\hat{A}^j
      -\hat{\partial}_j\hat{\partial}_j\hat{A}^i=&\frac{\beta_\star}{\epsilon\sqrt{\sigma}}\left(
      -\alpha\hat{\partial}_i\hat{\partial}_0\hat{\phi}+
      \epsilon_m\hat{n}\hat{u}^i\right).
\end{align}
The right hand side is regarded as the source terms which generate magnetic field and, hence, can not be much larger than the left hand side. In contrast to the Poisson equation \eqref{Poisson}, we allow for externally-applied magnetic field, which can exist ($A^i\ne0$) even when the right hand side is small or zero (that is vacuum magnetic field). Thus, we should consider only the following regime;
\begin{align}
\mbox{Balance 3: } \frac{\beta_\star}{\epsilon\sqrt{\sigma}}\max(\alpha,\epsilon_m)\le1. \label{balance3}    
\end{align}
Again, this inequality ``$\le$'' actually means ``$\lesssim$'' because this is a relation among the scale parameters.

Next, to estimate the orders of $f$ and $\tilde{f}$, let us normalize $N_+$ and $N_-$ as follows
\begin{align}
\hat{N}_\pm^2=\frac{N_\pm^2}{n_\star^2}
      =&\hat{n}^2(1-\beta_\star^2|\hat{\bm{v}}|^2)
      \pm2\frac{m_\mp}{m}\hat{n}\left(\alpha\hat{\tilde{n}}-\beta_\star^2\epsilon_m\hat{\bm{v}}\cdot\hat{n}\hat{\bm{u}}\right)\nonumber\\
      &+\frac{m_\mp^2}{m^2}\left(\alpha^2\hat{\tilde{n}}^2-\beta_\star^2\epsilon_m^2|\hat{n}\hat{\bm{u}}|^2\right).\label{N}
\end{align}
The first term on the right hand side is $O(1)$. Since we are interested in the case of $\alpha,\epsilon_m\ll1$ and the inequalities $m_\mp/m\le1$ and $\beta_\star<1$ always hold, the second and third terms on the right hand side are of small order; $\max(\alpha,\beta_\star^2\epsilon_m)\ll1$. As a loose assumption, we consider the situation where
\begin{align}
    \mbox{Balance 4: }\eta:=\max(\alpha,\beta_\star^2\epsilon_m)\le1,\label{balance4}
\end{align}
holds. Namely, we give up applying the MHD approximation when $\eta\gg1$.
Assuming \eqref{balance4}, we obtain the estimates, $n_\star f=O(1)$ and $n_\star\tilde{f}=O(\eta)$, and hence normalize them by $\hat{f}=n_\star f$ and $\hat{\tilde{f}}=n_\star\tilde{f}/\eta$. More explicitly, when $\eta\ll1$, the leading-order terms are calculated by series expansion as follows
\begin{align}
      \hat{f}=&\frac{\gamma}{\hat{n}}
      + \frac{\gamma^3}{2 \hat{n}}
      \mu^2(3\gamma^2\Lambda_a^2  - \Lambda_b)
      +O(\alpha^3,\beta_\star^2\epsilon_m^3),\label{f}\\
      \eta\hat{\tilde{f}}=&-\frac{\gamma^2}{\hat{n}}\Lambda_a+O(\alpha^2,\beta_\star^2\epsilon_m^2),
\end{align}
where
\begin{align}
\gamma=&\frac{1}{\sqrt{1-\beta_\star^2|\hat{\bm{v}}|^2}},\\
\Lambda_a:=&\alpha\frac{\hat{\tilde{n}}}{\hat{n}}-\beta_\star^2\epsilon_m\hat{\bm{v}}\cdot\hat{\bm{u}}=O(\alpha,\beta_\star^2\epsilon_m),\\
\Lambda_b:=&\alpha^2\frac{\hat{\tilde{n}}^2}{\hat{n}^2}-\beta_\star^2\epsilon_m^2|\hat{\bm{u}}|^2=O(\alpha^2,\beta_\star^2\epsilon_m^2).
\end{align}
Here, we emphasize that the first order terms in $\alpha$ and $\epsilon_m$ are vacant in the series expansion of $\hat{f}$, which turns out to be important later. 

It should be also remarked that we exclude the strongly-relativistic situation such as $\beta_\star|\hat{\bm{v}}|=|\bm{v}|/c=0.9999$, in which the Lorenz factor $\gamma$ becomes much greater than $1$ and our estimate $\hat{f}=O(1)$ is no longer valid. This means a breakdown of the assumed balance \footnote{If $x=O(1)$, the function $f(x)=1/\sqrt{1-x^2}$ is estimated as $O(1)$ in scale analysis. But, only the neighborhood of $x=1$ should be treated separately as an exceptional case due to singularity. For example, the method of matched asymptotic expansion is necessary for this kind of problems.} and strongly relativistic flow regions must be treated separately using a different normalization.
For example, we suggest that all the equations should be Lorenz-transformed to the inertia frame moving with the flow speed $0.9999c$ so that the Lorenz factor becomes $\gamma=O(1)$.

Now, the equations \eqref{mass_flux}, \eqref{current}, \eqref{par_Q} and \eqref{par_K} are normalized as follows
\begin{align}
\label{flux_or}
&\hat{\partial_0} \ \Bigl[ 
 \hat{f} \hat{Q^i} \hat{Q^0}
+ \mu^2 \eta \hat{\tilde{f}}\left(\alpha  \hat{Q^i} \hat{K^0}  + \epsilon_m \hat{K^i} \hat{Q^0}\right)
+ \mu^2 \alpha \epsilon_m \hat{f^\prime} \hat{K^i} \hat{K^0}\Bigr] \notag \\
& + \hat{\partial_j} \Bigl[
\hat{f}\hat{Q^i} \hat{Q^j}
+ \mu^2 \eta {\epsilon_m} \hat{\tilde{f}} ( \hat{Q^i} \hat{K^j} + \hat{Q^j} \hat{K^i} ) 
+ \mu^2 {\epsilon_m}^2 \hat{f^\prime} \hat{K^i} \hat{K^j}
\Bigr]\notag \\
&- \frac{\alpha^2}{\beta_\star^2 \epsilon^2} \hat{K^0} \hat{F}^i {}_0 
- \frac{\epsilon_m}{\beta_\star \epsilon} \sqrt{\sigma} \hat{K^j} \hat{F}^i {}_j = 0 , 
\end{align}
\begin{align}
\label{current_or}
&\hat{\partial_0} \ \Bigl[
\mu^2 \eta  \epsilon_m \hat{\tilde{f}} \hat{Q^i} \hat{Q^0}
+ \mu^2\epsilon_m  \hat{f^\prime}\left(  \alpha  \hat{Q^i} \hat{K^0}
            + \epsilon_m \hat{K^i} \hat{Q^0} \right)
\notag \\
& \hspace{30pt} - \mu^2 {\epsilon_m}^2 \alpha( \tilde{\mu} \hat{f^\prime} - \mu^2 \eta \hat{\tilde{f}} )  \hat{K^i} \hat{K^0} \Bigr] \notag \\ 
&+ \hat{\partial_j} \Bigl[
\mu^2 \eta \epsilon_m \hat{\tilde{f}} \hat{Q^i} \hat{Q^j}
+  \mu^2 {\epsilon_m}^2 \hat{f^\prime} ( \hat{Q^i} \hat{K^j} + \hat{Q^j} \hat{K^i} ) \notag \\
& \hspace{30pt}  - \mu^2 {\epsilon_m}^3 ( \tilde{\mu} \hat{f^\prime} - \mu^2 \eta \hat{\tilde{f}} )  \hat{K^i} \hat{K^j}
\Bigr] \notag \\
&- \frac{\alpha \epsilon_m }{\beta_\star^2 \epsilon^2}  (\hat{Q^0}- \tilde{\mu} \alpha \hat{K^0} )  \hat{F}^i {}_0 
 - \frac{\epsilon_m}{\beta_\star \epsilon}\sqrt{\sigma}
\qty(\hat{Q^j}- \tilde{\mu} \epsilon_m \hat{K^j} ) \hat{F}^i {}_j  = 0, 
\end{align}
\begin{eqnarray}
\label{par_Q_or}
\hat{\partial_0} \hat{Q^0} + \hat{\partial_i} \hat{Q^i} &=& 0, \\
\label{par_K_or}
\frac{\alpha}{\epsilon_m} \hat{\partial_0} \hat{K^0} + \hat{\partial_i} \hat{K^i} &=& 0, 
\end{eqnarray}
where the normalized electric field $\hat{E}^i$ is give by
\begin{align}
    \hat{E}^i=\hat{F}^i{}_0=\frac{cL_\star}{\phi_\star}F^i{}_0=-\hat{\partial}_i\hat{\phi}-\frac{\beta_\star\epsilon\sqrt{\sigma}}{\alpha}\hat{\partial}_0\hat{A}^i.
\end{align}

\subsection{Imposition of MHD balance \label{sec: reduce}}

The MHD approximation is understood as the reduction to a single-fluid model, satisfying
\begin{align}
    \alpha\ll1\quad\mbox{and}\quad\epsilon_m\ll1.
\end{align}
If they were not satisfied, we would have to solve the two-fluid equations as they are. However, in the limit of $\alpha,\epsilon_m\rightarrow0$ (then $\eta\rightarrow0$), many terms in \eqref{flux_or} are negligible and ultimately \eqref{flux_or} becomes the equation of motion for neutral fluid. Since this simple limit is not interesting, we assume that the electromagnetic ($\bm{J}\times\bm{B}$) force, which is $\hat{K^j} \hat{F}^i {}_j$ in \eqref{flux_or}, is not negligible but dominant. Namely, the flow $\bm{v}$ is dominantly accelerated by this term due to
\begin{align}
    \mbox{Balance 5: } \frac{\epsilon_m}{\beta_\star\epsilon}\sqrt{\sigma}=1.\label{MHD balance}
\end{align}
The another meaning of this balance can be understood by defining a representative cyclotron frequency as
\begin{align}
 \omega_{c\star}=  \frac{eB_\star}{m}= \frac{c}{ L_\star} \frac{\sqrt{ \sigma}}{\epsilon}.   
\end{align}
Then, the balance 5 indicates that
\begin{align}
    \epsilon_m=\frac{1/\omega_{c\star}}{T_\star}
\end{align}
is small if this frequency is faster than the time scale $T_\star$ of flow dynamics.
 It is well known that the two-fluid equations generally encompass ion's and electron's cyclotron motions. By taking the limit of $\epsilon_m\rightarrow0$ while keeping the balance 5, we can eliminate these fast motions from the flow dynamics. 
We also remark that the $\bm{v}\times\bm{B}$ term (which is $\hat{Q}^j\hat{F}^i{}_j$) in Ohm's law \eqref{current_or} becomes of order $1$ due to \eqref{MHD balance}.

Although the balance 5 is assumed in this work, we do not claim here that Hall or extended MHD must always satisfy $\epsilon_m=\omega_{c\star}^{-1}/T_\star\ll1$. 
In electron-ion plasmas, $\omega_{c\star}$ is approximately the ion cyclotron frequency $\omega_{ci}$, and the frequency of the whistler wave is known to be higher than $\omega_{ci}$, i.e. $\epsilon_m>1$. As will be shown later in Sec.~\ref{sec_reduction} and \ref{sec_range}, the electron inertia effect ($\epsilon_I^2$) can be neglected due to the smallness of the mass ratio $m_-/m_+$. If we take the limit of $m_-/m_+ \rightarrow 0$, the Hall and extended MHD would be applicable even if $\epsilon_m>1$.

\begin{figure}
    \centering
    \includegraphics[width=\linewidth]{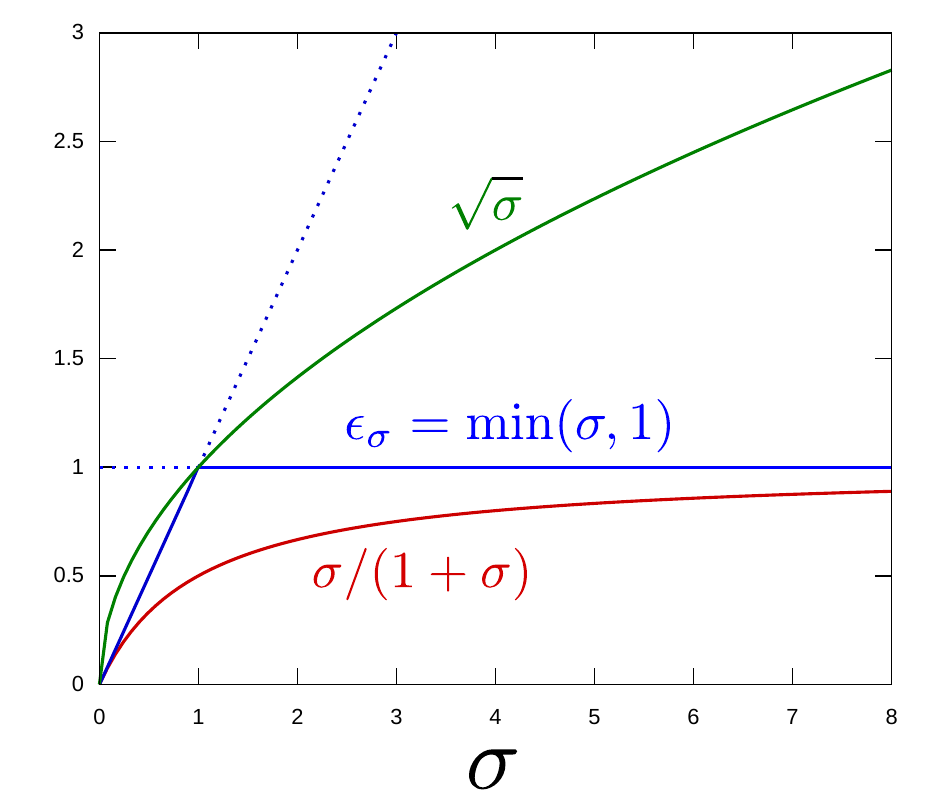}
    \caption{Plots of $\epsilon_\sigma$, $\sqrt{\sigma}$ and $\sigma/(1+\sigma)$}
    \label{fig:epsilon_sigma}
\end{figure}

On the other hand, the terms involving the electric field $\hat{F}^i{}_0$ in \eqref{flux_or} and \eqref{current_or} are, respectively, written as
\begin{align}
\frac{\alpha^2}{\beta_\star^2 \epsilon^2} \hat{K^0}\hat{F}^i{}_0
    =&
    \frac{1}{\sigma}\frac{\alpha^2}{\epsilon_m^2} \hat{K^0}\left[
    -\hat{\partial}_i\hat{\phi}-\sigma\frac{\epsilon_m}{\alpha}\hat{\partial}_0\hat{A}^i
    \right], \label{electric force1}
\end{align}
and
\begin{align}
    &\frac{\alpha \epsilon_m }{\beta_\star^2 \epsilon^2}  \qty(\hat{Q^0}- \tilde{\mu} \alpha \hat{K^0} )  \hat{F}^i {}_0\nonumber\\
    =&\frac{1}{\sigma}\frac{\alpha}{\epsilon_m}  \qty(\hat{Q^0}- \tilde{\mu} \alpha \hat{K^0} )\left[
    -\hat{\partial}_i\hat{\phi}-\sigma\frac{\epsilon_m}{\alpha}\hat{\partial}_0\hat{A}^i
    \right],\label{electric force2}
\end{align}
using the balance 5. In the limit of $\epsilon_m\rightarrow0$ or $\sigma\rightarrow0$, only the electrostatic force term ($\hat{\partial}_i\hat{\phi}$) gets too large $1/(\sigma\epsilon_m)\rightarrow\infty$ to balance with other terms. This implies that the existence of very fast plasma oscillation breaks down the assumed balance totally. To maintain the balance consistently, the charge separation $\alpha$ must be small enough that {\it all terms} in \eqref{electric force1} and \eqref{electric force2} are equal or less than the order $1$, which requires $\alpha\le\sqrt{\sigma}\epsilon_m$, $\alpha\le\sigma\epsilon_m$ and $\alpha\le\epsilon_m$. To consider the most general situation satisfying all of them, we assume
\begin{align}
    \mbox{Balance 6: } \alpha=\epsilon_m\epsilon_\sigma,\label{MHD balance2}
\end{align}
where $\epsilon_\sigma$ is the abbreviation of
\begin{align}
    \epsilon_\sigma:=\min\left(\sigma,1\right).
\end{align}
This is the last balance that we impose to derive MHD models. The magnitude of $\alpha$ is now determined by other scale parameters. The meaning of this balance is again understood by introducing a representative plasma frequency as
\begin{eqnarray}
\omega_{p\star}:=  c \sqrt{ \frac{e^2 \mu_0 n_\star}{m}} = \frac{1}{L_\star}\frac{c}{\epsilon}.
\end{eqnarray}
Since $\epsilon_\sigma\le \sqrt{\sigma}$ holds mathematically (see Fig.~\ref{fig:epsilon_sigma}), the balance 6 leads to
\begin{align}
    \alpha\le\frac{1/\omega_{p\star}}{T_\star}=\beta_\star\epsilon=\epsilon_m\sqrt{\sigma}.
\end{align}
Therefore, when the plasma frequency is much faster than the time scale of the flow dynamics ($\omega_{p\star}^{-1}/T_\star\rightarrow0$), the balance 6 requires $\alpha$ to be small ($\alpha\rightarrow0$), which diminishes the fast plasma oscillation. Note that $\omega_{p\star}^{-1}/T_\star$ is not always a small number when $\sigma$ is much greater than $1$. The balance 6 requires smallness of $\alpha$ more strictly than the condition $\alpha\le\omega_{p\star}^{-1}/T_\star$ when $\sigma\ge1$.

At this point, we summarize the situation where all the balances $1,2,\dots,6$ are imposed together. Given the balances 5 and 6, the balance 3 can be reduced to
\begin{align}
    \mbox{Balance 3': }\frac{\beta_\star^2}{\sigma}\le1,
\end{align}
Therefore, the situation can be divided into the two cases, $\beta_\star^2\le\sigma\le1$ or $1\le\sigma$. 
In either case, the balance 4 is simply rewritten as
\begin{align}
    \mbox{Balance 4': }&\eta=\epsilon_m\epsilon_\sigma\le1.
\end{align}

By omitting the hat symbol $\hat{\phantom{a}}$ in what follows, the normalized equations are summarized as follows
\begin{align}
&\partial_0 \ \Bigl [
 f Q^iQ^0
+ \epsilon_I^2\epsilon_\sigma \tilde{f}\left(\epsilon_\sigma  Q^i K^0  + K^iQ^0\right)
+ \epsilon_\sigma \epsilon_I^2 f^\prime K^i K^0\ \Bigr ] \notag \\
&+ \partial_j \ \Bigl [
f  Q^i Q^j
+ \epsilon_I^2\epsilon_\sigma \tilde{f} \qty( Q^i K^j  +  Q^j K^i ) + \epsilon_I^2 f^\prime K^i K^j \
\Bigr ] \notag \\
&
-\epsilon_\sigma K^0\left[
    -\frac{\epsilon_\sigma}{\sigma}\partial_i \phi-\partial_0 A^i
    \right]
- K^j F^i {}_j = 0,\label{Momentum1} 
\end{align}
\begin{align}
&\epsilon_I^2 \partial_0 \ \Bigl [
\epsilon_\sigma   \tilde{f} Q^i Q^0
+  f^\prime\left(  \epsilon_\sigma  Q^i K^0  +  K^i Q^0 \right)
\notag \\
& \hspace{100pt} - \epsilon_\sigma\qty(\epsilon_H f^\prime - \epsilon_\sigma\epsilon_I^2 \tilde{f} )  K^i K^0 \ \Bigr ] \notag \\ 
& + \epsilon_I^2 \partial_j \ \Bigl [
 \epsilon_\sigma \tilde{f} Q^i Q^j
+   f^\prime \qty( Q^i K^j + Q^j K^i ) \notag \\
& \hspace{120pt}  -  \qty( \epsilon_H f^\prime - \epsilon_\sigma\epsilon_I^2 \tilde{f} )  K^i K^j
\Bigr ] \notag \\
&
-\qty(Q^0- \epsilon_H\epsilon_\sigma K^0 )\left[
    -\frac{\epsilon_\sigma}{\sigma} \partial_i\phi-\partial_0A^i
    \right]\notag\\
&    - \qty(Q^j- \epsilon_H K^j ) F^i {}_j  = 0, \label{Momentum2}
\end{align}
\begin{align}
\partial_0 Q^0 + \partial_i Q^i &= 0, \\
\epsilon_\sigma\partial_0K^0 + \partial_i K^i &=0, 
\end{align}
\begin{align}
-\partial_j\partial_j\phi=\tilde{n},\\
      \beta_\star^2\partial_0\partial_0A^i
      +\partial_i\partial_jA^j
      -\partial_j\partial_jA^i=&\frac{\beta_\star^2}{\sigma}\left(
      -\epsilon_\sigma\partial_i\partial_0\phi+
       nu^i\right),\label{AM2}
\end{align}
where 
\begin{align}
    f'=f-\epsilon_H\epsilon_\sigma \tilde{f}.
\end{align}
We have introduced
\begin{align}
    \epsilon_H := \tilde{\mu} \epsilon_m,\\
    \epsilon_I := \mu \epsilon_m,
\end{align}
because $\epsilon_m$ appears only in these forms.
By noting $\Lambda_a=O(\epsilon_m \epsilon_\sigma)$ and $\Lambda_b=O(\epsilon_m^2\epsilon_\sigma)$, the estimate \eqref{f} becomes
\begin{align}
      f=&\frac{1}{n\sqrt{1-\beta_\star^2|\bm{v}|^2}}+O(\epsilon_\sigma\epsilon_I^2).\label{f2}
\end{align}

Due to the balances 5 and 6, the number of non-dimensional parameters (namely, the dimension of the parameter space) has been reduced to three; $\epsilon_m,\sigma,\beta_\star$. However, the governing equations are still equivalent to the two-fluid equations.
In the following sections, we derive reduced models by taking the specific limit of $\epsilon_m,\sigma,\beta_\star$. 

\section{Reduction to various MHD models}\label{sec_reduction}

\subsection{Vacuum limit $\beta_\star^2/\sigma\rightarrow0$}

The limit $\beta_\star^2/\sigma\rightarrow0$ corresponds to vacuum state since the right hand side of the Ampere-Maxwell equation \eqref{AM2} vanishes. This limit may be uninteresting because the plasma current is too small to disturb the vacuum magnetic field. For example, the limit of large magnetization parameter $\sigma\rightarrow\infty$ inevitably results in this vacuum state (due to $\beta_\star \le1$)

As another high magnetization limit $\sigma\rightarrow\infty$, the force-free approximation is widely used~\cite{komissarov_time-dependent_2002}. When the current is divided into parallel ($J_\parallel$) and perpendicular ($\bm{J}_\perp$) components to the magnetic field, $J_\parallel$ remains finite while $\bm{J}_\perp=0$ (i.e., $\bm{J}\times\bm{B}=0$) in the force-free state. In this study, we set $n_\star u_\star$ as the single representative scale of the current, and no distinction is made between the parallel and perpendicular components.
To obtain the force-free state, only the parallel component should diverge ($u_{\parallel\star}\rightarrow\infty$) to keep the parallel current finite in the vacuum limit $n_\star\rightarrow0$. 
Although the force-free model will be reproduced later in Sec.~\ref{sec: HR}, the existence of the "extended" force-free model is non-trivial and will be the subject of future work.

According to the dispersion relation of RMHD, the relativistic Alfv\'en velocity $v_{A\star}$ is well-known as
 \begin{align}
     v_{A*} = \frac{ B_\star c}{\sqrt{\mu_0 m n_\star c^2 + B_\star^2}}=\sqrt{\frac{\sigma}{1+\sigma}}c.
 \end{align}
When $\sigma\rightarrow\infty$ (i.e., strong magnetic field or low density limit), the displacement current becomes dominant and the Alfv\'en wave turns into the electromagnetic wave in vacuum. 

The interaction between plasma motion and electromagnetic field is most active in the situation $\beta_\star^2/\sigma\simeq1$. Therefore, the Alfv\'en ordering $\beta_\star^2/\sigma=1$ is conventionally employed in (non-relativistic) MHD focusing on only this situation, which is admissible as far as $\sigma\le1$.
It is also interesting to note that the scale of $\epsilon_\sigma$ is similar to $v_{A\star}^2/c^2=\sigma/(1+\sigma)$.

\subsection{Single-fluid limit $\epsilon_m\rightarrow0$}

In the limit of $\epsilon_m\rightarrow0$ (then $\epsilon_H,\epsilon_I\rightarrow0$), a lot of terms can be neglected in \eqref{Momentum1} and \eqref{Momentum2} as follows
\begin{align}
&\partial_0( \gamma n v^i) 
+ \partial_j(\gamma n  v^i v^j) \notag \\
&-\epsilon_\sigma \tilde{n}\left(
    -\frac{\epsilon_\sigma}{\sigma}\partial_i \phi-\partial_0 A^i
    \right)
- nu^j F^i {}_j = 0 , \\
& 
-n\left(
    -\frac{\epsilon_\sigma }{\sigma} \partial_i\phi-\partial_0A^i
    \right)
    - nv^j F^i {}_j  = 0.
\end{align}
These are the well-known RMHD equations, where
the expression of $f$ has been simplified into 
\begin{align}
    f=\frac{\gamma}{n}=\frac{1}{n\sqrt{1-\beta_\star^2|\bm{v}|^2}}.
\end{align}
 and $\tilde{f}$ is completely neglected as if the ``proper charge neutrality'' $N_+=N_-$ holds. In fact, there exists small-order charge separation $\tilde{n}\ne0$ (with $\nabla\cdot\bm{J}\ne0$) and the associated electrostatic force takes part in the dominant balance.
We will discuss more about the RMHD equations in Sec.~\ref{sec: HR}. 

\subsection{Non-relativistic limit $\sigma\rightarrow0$}

Here, we consider the limit of small $\sigma$.
Because of the balance 3', the limit of $\sigma\rightarrow0$ forces $\beta_\star^2\rightarrow0$ as well (and $\epsilon_\sigma\rightarrow0$). Therefore, let us take these non-relativistic limits while keeping
\begin{align}
    \frac{\beta_\star^2}{\sigma}\rightarrow\frac{v_\star^2}{v_{A\star}^2}=\mbox{const.}(\le1).
\end{align}
Then, the equations \eqref{Momentum1}, \eqref{Momentum2} and \eqref{AM2} are reduced to the extended MHD (XMHD) equations,
\begin{align}
&\partial_0 \left(nv^i\right) + \partial_j \left(
nv^i v^j
+  \epsilon_I^2 nu^i u^j
\right)
- nu^j F^i {}_j = 0,\\
&\epsilon_I^2\partial_0(nu^i)
+\epsilon_I^2 \partial_j \left[ 
  n( v^i u^j + v^j u^i )  -   \epsilon_H n  u^i u^j
\right] \notag \\
&\hspace{10pt} -n\left(
    -\partial_i\phi-\partial_0A^i
    \right)
    - n(v^j- \epsilon_H u^j ) F^i {}_j  = 0, \label{Ohm_XMHD}\\
&     \partial_i\partial_jA^j
      -\partial_j\partial_jA^i=\frac{v_\star^2}{v_{A\star}^2}
       nu^i.
\end{align}
where $f=1/n$ has been substituted. Since $v_\star^2/v_{A\star}^2\rightarrow0$ is again the uninteresting vacuum limit, it is conventional to use the Alfv\'en ordering $v_\star=v_{A\star}$.

The equation \eqref{Ohm_XMHD} is the generalized Ohm's law. Recall that we neglected resistivity and viscosity at the beginning, assuming that they are not dominant. The inclusion of these terms with large Lundquist and Reynolds numbers does not affect the balances we have discussed so far.

Note that the displacement current in \eqref{AM2} has been neglected and hence
$\nabla\times\bm{B}=(v_\star^2/v_{A\star}^2)\bm{J}$ is a constraint; we can eliminate $nu^i$ (or $\bm{J}$) using this relation. The generalized Ohm's law \eqref{Ohm_XMHD} is regarded as the evolution equation for $A^i$, where $\phi$ is determined such that the Coulomb gauge $\partial_iA^i=0$ holds.  Therefore, the XMHD equations constitute a dynamical system of 7 fields $(n,v^i,A^i)$ with one constraint $\partial_iA^i=0$. Since $\nabla\cdot\bm{J}=\partial_i(nu^i)=0$, the so-called ``quasi-neutrality condition'' holds as if $\tilde{n}=0$. In fact, small charge separation $\tilde{n}=-\partial_j\partial_j\phi\ne0$ exists although it no longer appears explicitly in the XMHD equations. For small $\sigma\ll1$, the balance 6 ($\alpha=\epsilon_m\sigma$) requests $\alpha$ to be further smaller than $\epsilon_m$. The charge conservation law is indeed reduced to $\nabla\cdot\bm{J}=0$ in the limit $\sigma\rightarrow0$.

The electron-inertia effect is manifested by the terms with $\epsilon_I^2$, which is the second-order of $\epsilon_m$. 
If we neglect only $\epsilon_I^2$, the Hall MHD (HMHD) equations are reproduced. If we neglect $\epsilon_H$ too or simply take the limit of $\epsilon_m\rightarrow0$, the MHD equations are finally obtained.

\subsection{Relativistic Hall MHD model}

Now, we are positioned to search for the other MHD models which include both the relativistic and two-fluid effects.
To derive a reduced model, we still need to assume the smallness of $\epsilon_m$ but should not neglect it completely.
An approximation that comes to mind immediately is to neglect $O(\epsilon_m^2)$, namely, the electron-inertia effect $\epsilon_I^2$ only.
The resultant equations deserve to be called relativistic Hall MHD (RHMHD). It is remarkable that $f$ in \eqref{f2} includes no additional term due to the Hall effect, $O(\epsilon_H)$ or $O(\epsilon_m)$. By neglecting $\epsilon_I^2$, the proper charge neutrality $f=\gamma/n$ still holds approximately and $\tilde{f}$ vanieshes.

Therefore, in RHMHD, \eqref{Momentum1} and \eqref{Momentum2} are reduced to
\begin{align}
&\partial_0 \left(
 \gamma n v^i \right)  + \partial_j\left(
\gamma nv^i v^j
\right)  \notag \\
& 
-\epsilon_\sigma \tilde{n}\left(
    -\frac{\epsilon_\sigma}{\sigma}\partial_i \phi-\partial_0 A^i
    \right)
- nu^j F^i {}_j = 0,\\ 
&
-\left(n- \epsilon_H\epsilon_\sigma \tilde{n} \right)\left(
    -\frac{\epsilon_\sigma}{\sigma} \partial_i\phi-\partial_0A^i
    \right) \notag \\
&\hspace{10pt} - n(v^j- \epsilon_H u^j ) F^i {}_j  = 0.
\end{align}
The terms including $\epsilon_H$ are the difference from RMHD.

If we further assume the smallness of $\sigma\ll1$ additionally, we can also neglect the term of $O(\epsilon_H\epsilon_\sigma)$ in Ohm's law. 
Although it is just a minor reduction, let us call it weakly-relativistic Hall MHD (W-RHMHD).

\subsection{Weakly-relativistic XMHD model}

If one wants to allow for both the electron-inertia and relativistic effects, it is difficult to derive a reduced model from the two-fluid equations. One option is to neglect $O(\epsilon_I^2\sigma)$ by assuming the sufficient smallness of both $\epsilon_m$ and $\sigma(\ll1)$. 
Then, the proper charge neutrality holds again and we can derive similar equations to XMHD.
\begin{align}
&\partial_0\left(\gamma n v^i\right)
 + \partial_j\left(
\gamma nv^i v^j
 + \epsilon_I^2 \gamma nu^i u^j\right) \notag \\
&\hspace{10pt}  
-\sigma \tilde{n}\left(
    -\partial_i \phi-\partial_0 A^i
    \right)
- nu^j F^i {}_j = 0, 
\end{align}
\begin{align}
&\epsilon_I^2 \partial_0 \left(\gamma n u^i \right)
 + \epsilon_I^2 \partial_j \ \left[ 
  \gamma n\left( v^i u^j + v^j u^i \right) -\epsilon_H \gamma n u^iu^j
\right] \notag \\
&\hspace{10pt}  
-\left(n- \epsilon_H\sigma \tilde{n}\right)\left(
    -\partial_i \phi-\partial_0 A^i
    \right)
    - n\left(v^j- \epsilon_H u^j \right) F^i {}_j  = 0.
\end{align}
We call this model weakly-relativistic XMHD because it is valid only when $\sigma$ and $\beta_\star^2$ are sufficiently smaller than $1$.

\section{Range of application\label{sec_range}}

Under the balances 1 to 6, there remain three non-dimensional parameters $(\epsilon_m,\sigma,\beta_\star)$, which are related to the three physical scales $(n_\star,v_\star,B_\star)$ of plasma (for fixed length scale $L_\star$). More rigorously speaking, $\tilde{\mu}$ and $\mu$ are two additional parameters which appear only in the forms, $\epsilon_H=\tilde{\mu}\epsilon_m$ and $\epsilon_I=\mu\epsilon_m$. Because of the inequalities $\tilde{\mu}\le1$ and $\mu\le1$, they do not alter the balances 1 to 6 but possibly make $\epsilon_H$ and $\epsilon_I$ further smaller than $\epsilon_m$. As the two typical examples, we will consider electron-ion (Hydrogen) plasma $m_-/m_+\simeq10^{-3}$ (for which $\mu\simeq10^{-3/2}=0.0316$ and $\tilde{\mu}\simeq1$) and electron-positron plasma $m_-/m_+=1$ (for which $\mu=0.5$ and $\tilde{\mu}=0$). 

Now, we carefully consider the magnitude of $\epsilon_m$, which obviously measures the impact of the two-fluid effect as we have seen in the previous section. According to the balance 5 and 3', it depends on the other parameters as follows
\begin{align}
    \epsilon_m=\frac{\beta_\star}{\sqrt{\sigma}}\epsilon=\frac{v_\star}{L_\star \omega_{c\star}}\quad\mbox{with}\quad \frac{\beta_\star}{\sqrt{\sigma}}\le1. \label{epsilon_m}
\end{align}
Here, we obtain $\epsilon=d_\star/L_\star$ by newly introducing a representative inertial length (or skin depth) $d_\star=c/\omega_{p\star}$. 
More specifically, the inertial lengths for positively ($+$) and negatively ($-$) charged gases are given by
\begin{align}
    d_\pm=\sqrt{\frac{m_\pm}{m_++m_-}}d_\star.
\end{align}
Indeed, $d_+$ and $d_-$ respectively correspond to ion's and electron's inertial lengths for electron-ion plasma.

In the case of non-relativistic limit ($\sigma\rightarrow0$ and $\beta_\star\rightarrow0$)  with application of the Alfv\'en ordering $\beta_\star^2/\sigma\rightarrow1$ ($v_\star=v_{A\star}$), we simply obtain $\epsilon_m=\epsilon$. Therefore,
\begin{align}
  \epsilon_H\simeq d_+/L_\star\quad\mbox{and}\quad  \epsilon_I\simeq d_-/L_\star (\simeq\sqrt{m_-/m_+}\epsilon_H),
\end{align}
for electron-ion plasma, and
\begin{align}
  \epsilon_H=0 \quad\mbox{and}\quad \epsilon_I=\sqrt{2}d_+/L_\star=\sqrt{2}d_-/L_\star,
\end{align}
 for electron-positron plasma. In this way, the Hall and electron-inertia effects are associated with the small-scales $d_+$ and $d_-$, where the density $n_\star$ is important because $d_\star$ is proportional to $1/\sqrt{n_\star}$ only. For sufficiently dense plasma $n_\star\rightarrow\infty$, the two-fluid effect becomes negligible $\epsilon\rightarrow0$. However, this is a consequence of applying the Alfv\'en ordering. Namely, $v_\star$ and $B_\star$ are not fixed independently but varied along with $n_\star$.

In general, it is interesting to note that $\epsilon_m$ does not originally depend on the density $n_\star$ but on the ratio  $v_\star/B_\star$ in \eqref{epsilon_m}. When $\sigma$ gets larger than $1$, the two-fluid effect $\epsilon_m$ gets smaller than $\epsilon$ by the factor $\beta_\star/\sqrt{\sigma}$.  
This tendency agrees with Kawazura~\cite{kawazura2017Hall} in which the ion skin depth is modified to shrink as the magnetic field strength increases relativistically. In the limit of $B_\star\rightarrow\infty$ with fixed $n_\star$ and $v_\star$, we find that the two-fluid effect becomes negligible and the use of RMHD is justified (although it tends to be almost vacuum plasma, $\beta_\star^2/\sigma\rightarrow0$). 

In the case of electron-positron plasma, the Hall effect vanishes identically, $\epsilon_H=0$. Then, the XMHD model includes only the electron-inertia effect, which is especially called Inertial MHD (IMHD). Similarly, we can obtain weakly-relativistic IMHD from weakly-relativistic XMHD when $\epsilon_H=0$. 

\begin{table}[h]
\begin{center}
\renewcommand\arraystretch{1.2}
\begin{tabular}{|c|l|l|}
\hline
 \textbf{Model} & \textbf{Included order} & \textbf{Neglected order} \\
\hline \hline
  MHD & & $O(\sigma)$, $O(\epsilon_H)$, $O(\epsilon_I^2)$ \\
  \hline
HMHD & $O(\epsilon_H)$  & $O(\sigma)$, $O(\epsilon_I^2)$ \\
   \hline
IMHD & $O(\epsilon_I^2)$  & $O(\sigma)$, $O(\epsilon_H)$  \\
 \hline
XMHD & $O(\epsilon_H)$, $O(\epsilon_I^2)$,  & $O(\sigma)$ \\
 \hline
 RMHD &  $O(\sigma)$ & $O(\epsilon_H)$, $O(\epsilon_I^2)$ \\
   \hline
 RHMHD &  $O(\sigma)$, $O(\epsilon_H)$  & $O(\epsilon_I^2)$ \\
     \hline
 W-RHMHD &  $O(\sigma)$, $O(\epsilon_H)$  & $O(\epsilon_I^2)$, $O(\epsilon_H\sigma)$ \\
  \hline
 W-RIMHD & $O(\sigma)$,  $O(\epsilon_I^2)$ & $O(\epsilon_H)$, $O(\epsilon_I^2\sigma)$ \\
    \hline
 W-RXMHD & $O(\sigma)$, $O(\epsilon_H)$, $O(\epsilon_I^2)$ & $O(\epsilon_I^2\sigma)$ \\
\hline
\end{tabular}
\caption{Classification of models (H = Hall, I = Inertial, X = eXtended, R = Relativistic, W-R = Weakly-Relativistic) }
\label{table: EI model}
\end{center}
\end{table}

All the models which we have presented so far are summarized in Table~\ref{table: EI model}.
All these reduced models need to neglect the order of
\begin{align}
    \epsilon_\sigma\epsilon_I^2=\frac{m_+m_-}{m^2}\beta_\star^2\frac{\epsilon_\sigma}{\sigma}\epsilon^2,
\end{align}
in common, which is necessary for $f=\gamma/n$ to hold approximately and to get rid of $\tilde{f}$. Then, either proper charge neutrality or quasi-neutrality holds. In other words, we have to solve the two-fluid equations directly if this $\epsilon_\sigma\epsilon_I^2$ is not sufficiently smaller than $1$.

\begin{figure*}[ht]
    \includegraphics[width=0.8\textwidth]{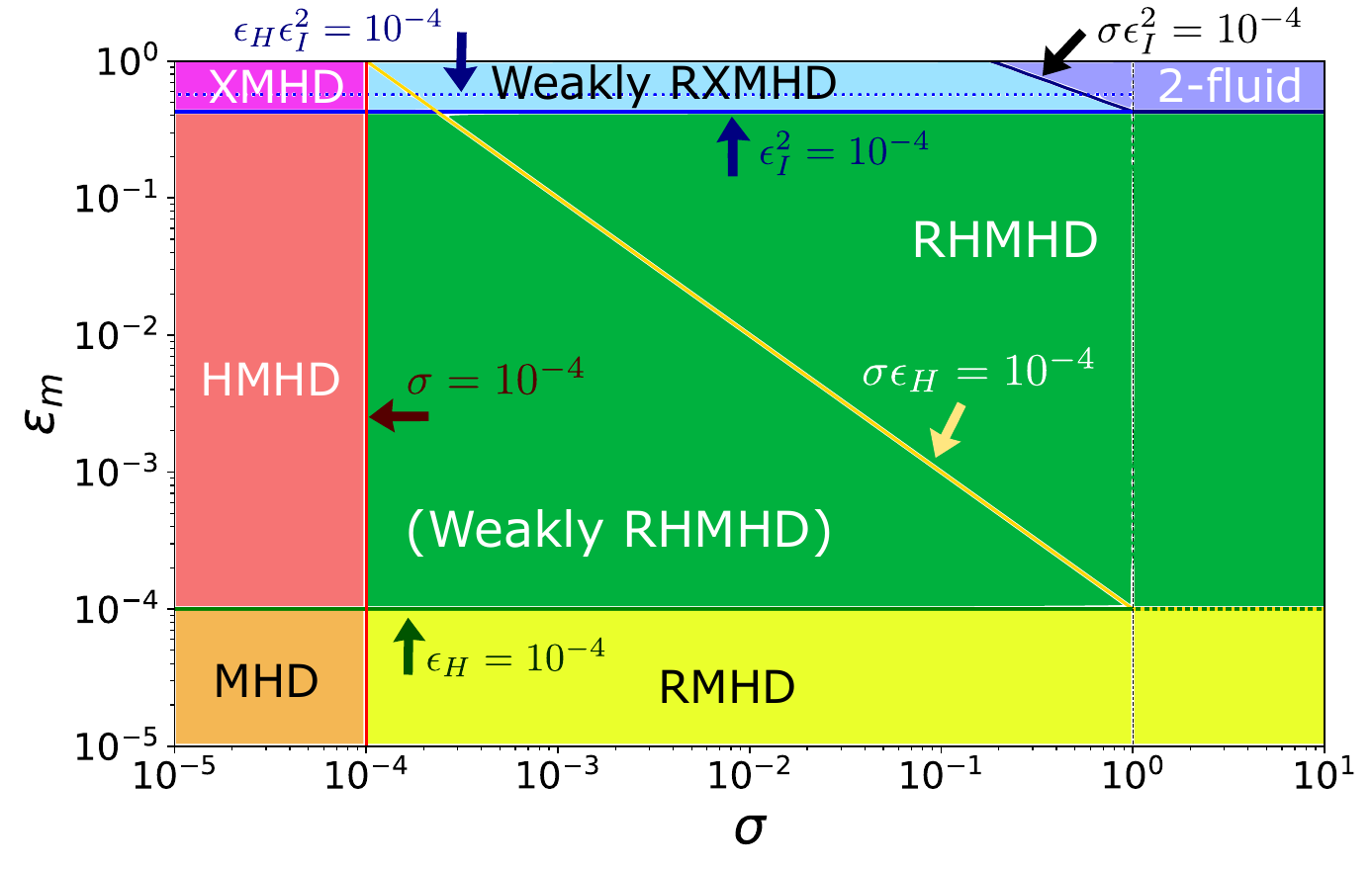}
    \caption{Electron-Ion Plasma ($\sigma$ vs. $\epsilon_m$)}
    \label{fig:ion_map1}
\end{figure*}

\begin{figure*}[ht]
    \includegraphics[width=0.8\textwidth]{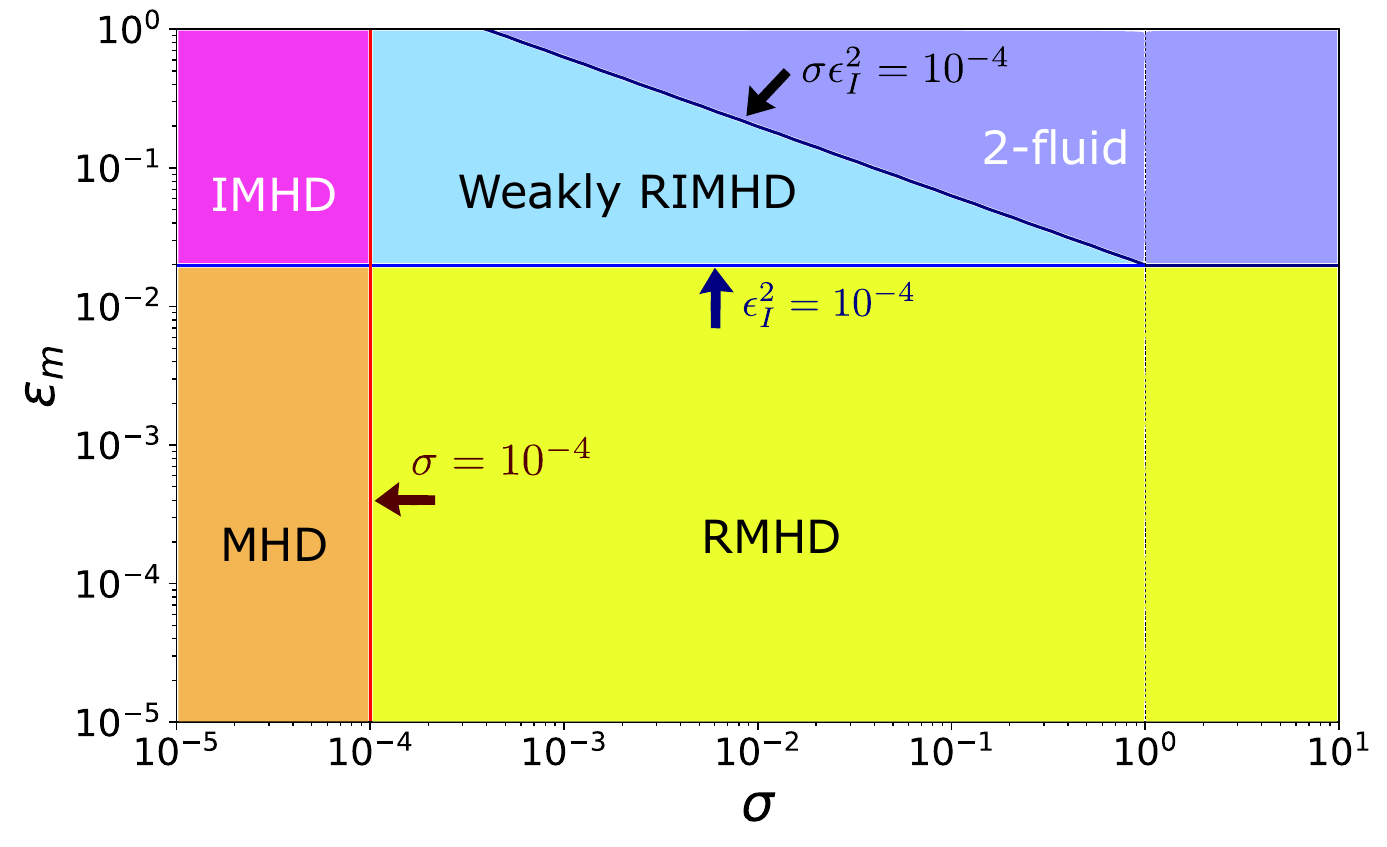}
    \caption{Electron-Positron Plasma ($\sigma$ vs. $\epsilon_m$)}
    \label{fig:positron_map1}
\end{figure*}

Since it is still difficult to imagine the applicable scope of each model, let us assume $10^{-4}$ as a clear threshold for example. Namely, the non-dimensional parameters (such as $\sigma$ and $\epsilon_H$) can be considered negligible if they are below $10^{-4}$. Otherwise, they are not neglected. Then, the lines such as  $\sigma=10^{-4}$ and $\epsilon_H=10^{-4}$ divide the parameter space $(\epsilon_m,\sigma,\beta_\star)$ into subspaces, in which a certain group of MHD models is applicable. Recall that $\beta_\star$ should satisfy $\beta_\star^2\le\sigma$ and it appears only in Ampere-Maxwell's equation \eqref{AM2}. As in Table~\ref{table: EI model}, the models are classified in terms of $\sigma$ and $\epsilon_m$ (regardless of $\beta_\star$), which are illustrated in Fig.~\ref{fig:ion_map1} for electron-ion plasma and in Fig.~\ref{fig:positron_map1} for electron-positron plasmas.
Due to the smallness of mass ratio $m_-/m_+\simeq10^{-3}$, the electron-inertia effect is readily neglected ($\epsilon_I^2<10^{-4}$) in the majority of cases for electron-ion plasma. But, we have to keep in mind that electron inertia may be important {\it locally} at singular point or layer, where the small-scale structure $L_\star\sim d_-$ emerges (as in the location where magnetic reconnection occurs).

\begin{figure*}[ht]
    \includegraphics[width=0.8\textwidth]{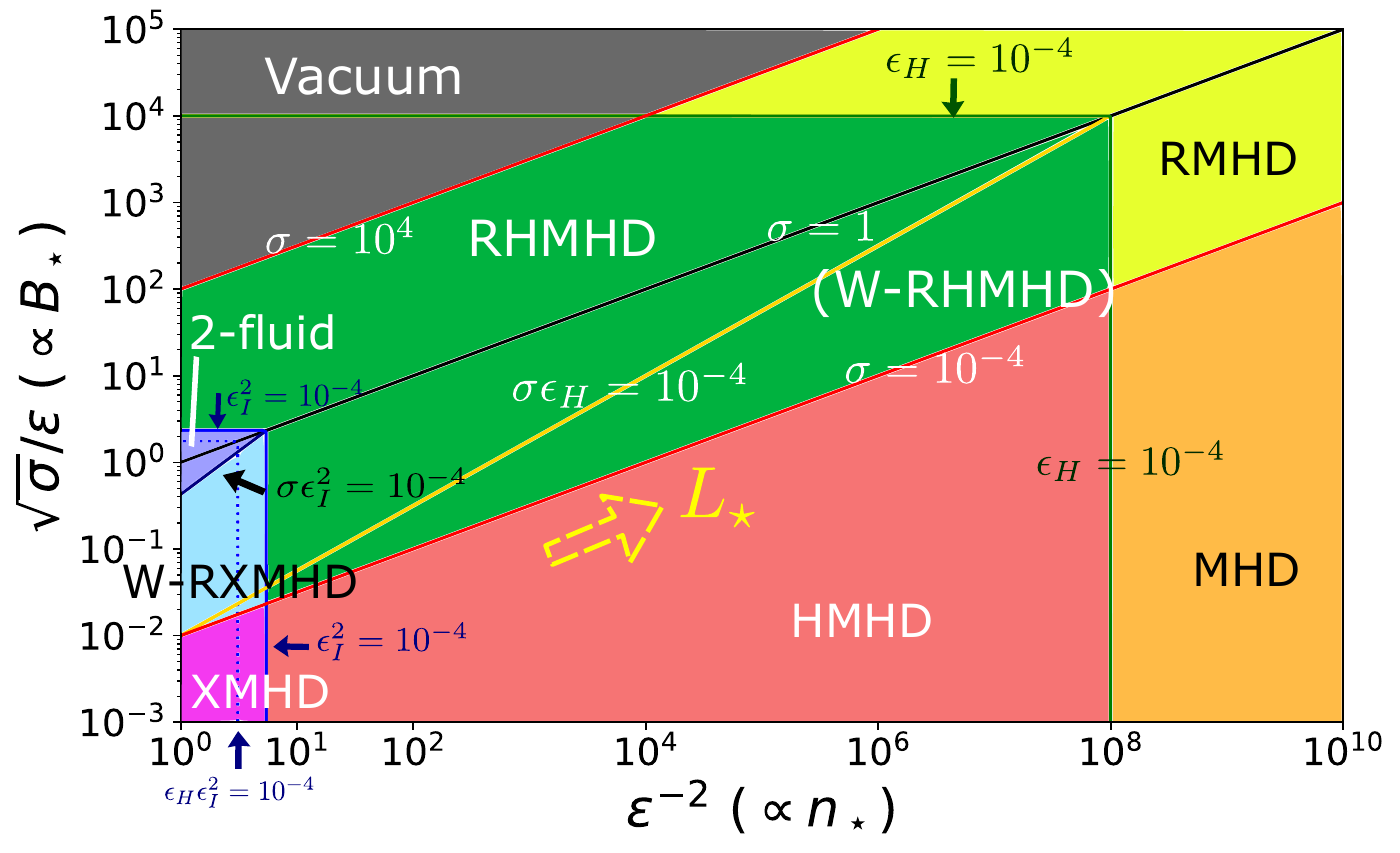}
    \caption{Electron-Ion Plasma ($n_\star$ vs. $B_\star$ vs. $L_\star$)}
    \label{fig:ion_map2}
\end{figure*}

\begin{figure*}[ht]
    \includegraphics[width=0.8\textwidth]{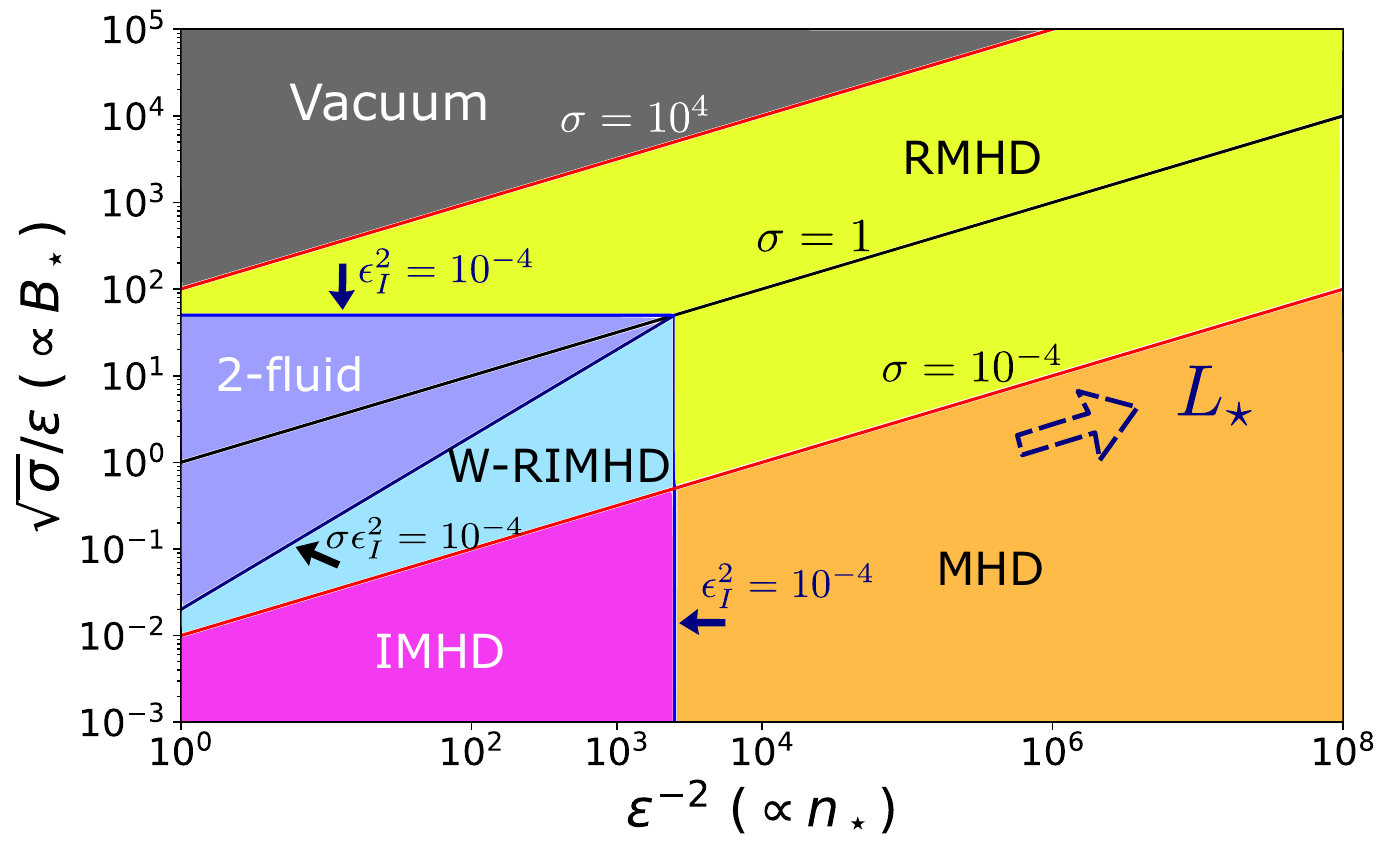}
    \caption{Electron-Positron Plasma ($n_\star$ vs. $B_\star$ vs. $L_\star$)}
    \label{fig:positron_map2}
\end{figure*}

Although Figs.~\ref{fig:ion_map1} and \ref{fig:positron_map1} look simple enough, let us further illustrate the application ranges in terms of $(n_\star,v_\star,B_\star)$.
Since the vacuum state $\beta_\star^2/\sigma\ll1$ is uninteresting, it is reasonable to fix $\beta_\star$ to the maximal value $\beta_\star=\sqrt{\epsilon_\sigma}$.
As shown in the plots of Fig.~\ref{fig:epsilon_sigma}, there is in fact no order difference between $\sqrt{\epsilon_\sigma}$ and $v_{A\star}/c=\sqrt{\sigma/(1+\sigma)}$
 in the scale analysis; $\sqrt{\epsilon_\sigma}\simeq v_{A\star}/c$. Therefore, we refer to 
 \begin{align}
  \beta_\star=\sqrt{\epsilon_\sigma}\simeq v_{A\star}/c   
 \end{align}
 as relativistic Alfv\'en ordering.
By imposing this relativistic Alfv\'en ordering on $\beta_\star$, we obtain $\epsilon_m=\epsilon\sqrt{\epsilon_\sigma/\sigma}$ and the two remaining parameters are chosen as
\begin{align}
    \epsilon^{-2}=\frac{\mu_0n_\star e^2L_\star^2}{m}\propto n_\star,\\
    \sqrt{\sigma}/\epsilon=\frac{eB_\star L_\star}{mc}\propto B_\star,
\end{align}
representing the scales of $(n_\star,B_\star)$ directly. Therefore, we can remap Fig.~\ref{fig:ion_map1} into Fig.~\ref{fig:ion_map2} and Fig.~\ref{fig:positron_map1} into Fig.~\ref{fig:positron_map2} on the 2d plane $(\epsilon^{-2},\sqrt{\sigma}/\epsilon)$.
The region of $\sigma>10^4$ is filled in gray because it is considered vacuum ($\beta_\star^2/\sigma<10^{-4}$). In the strong magnetic field limit $B_\star\rightarrow\infty$, we inevitably enter this vacuum regime but RMHD is still valid and no problem to keep using it. In the dense plasma limit $n_\star\rightarrow\infty$, we enter the conventional MHD regime. In this figure, the limit of large scale $L_\star\rightarrow\infty$ corresponds to the movement in the direction indicated by the fat arrow, which is parallel to the $\sigma=$const. line. In the triangle region indicated by "2-fluid", the charge neutrality approximation $\epsilon_\sigma\epsilon_I^2<10^{-4}$ is not satisfied. This region exists on the low density side $\epsilon^{-2}<10^4\mu^2$ and for intermediate strength of magnetic field. In the weak magnetic field limit $B_\star\rightarrow0$, we can apply the non-relativistic MHD models (such as XMHD, HMHD, IMHD, MHD). But, when magnetic field is weak such that $\sqrt{\sigma}/\epsilon<\sqrt{10^4\mu^2}$ holds and in the low density limit $n_\star\rightarrow0$, we have to solve the two-fluid equations without assuming charge neutrality. The limit of small scale $L_\star\rightarrow0$ also enters the "2-fluid" region eventually.

\section{Remarks on RHMHD}\label{sec: HR}

In comparison to RMHD, the RHMHD equations just have a few additional terms in Ohm's law due to the Hall effect. But, this difference is quite influential when solving these equations theoretically and numerically. 

In general, when a dynamical system $\partial_t u=F(u)$ for $u$ is solved numerically, the recurrence formula such as $u^{n+1}=u^n+F(u^n)\Delta t$ is iterated for time marching $t\rightarrow t+\Delta t$. To execute this iteration, the right hand side $F(u)$ must be calculated uniquely using the dynamical variable $u$. This is a fundamental requirement for the well-posedness of time-evolving system.

In 3d vector format, the RHMHD equations are composed of the evolution equations (that include time derivative $\partial_t$ of some quantity),
\begin{align}
\partial_tn=&-\nabla\cdot(n\bm{v}),\label{C}\\
\partial_t( n \gamma\bm{v})=&- \nabla\cdot(n \gamma \bm{v}\bm{v}) 
+\epsilon_\sigma \tilde{n}\bm{E}
+ \bm{J}\times\bm{B}, \label{MomentP}\\
    \beta_\star^2\partial_t\bm{E}
      =&\nabla\times\bm{B}-\frac{\beta_\star^2}{\sigma}\bm{J},\label{AM}\\
 \partial_t\bm{B}=&-\nabla\times\bm{E},\label{FM}
\end{align}
and the constraints,
\begin{align}
n(\bm{E}+ \bm{v}\times\bm{B}) - \epsilon_H(\epsilon_\sigma \tilde{n}\bm{E}
+\bm{J}\times\bm{B}) = 0, \label{Ohm}\\
\nabla\cdot\bm{E}=\frac{\epsilon_\sigma}{\sigma}\tilde{n},\\
\nabla\cdot\bm{B}=0.
\end{align}
A drastic change from the two-fluid equations is that the time derivative of the current $\bm{J}$ no longer exists in \eqref{Ohm} due to neglect of electron-inertia $\epsilon_I^2\rightarrow0$. Therefore, to calculate the right hand sides of the evolution equations, we have to determine (or eliminate) $\bm{J}$ using the other variables.

From Ohm's law \eqref{Ohm}, the electric field of the component parallel to the magnetic field must be zero ($\bm{E}\cdot\bm{B}=0$).
By combining \eqref{AM} and \eqref{FM}, we obtain
\begin{align}
      \partial_t(\bm{E}\times\bm{B})=&
      \nabla\cdot\left(\frac{\bm{B}\bm{B}}{\beta_\star^2}+\bm{E}\bm{E}\right)
      -\nabla\left(\frac{|\bm{B}|^2}{2\beta_\star^2}+\frac{|\bm{E}|^2}{2}\right)\nonumber\\
     & -\frac{1}{\sigma}(\epsilon_\sigma\tilde{n}\bm{E}+\bm{J}\times\bm{B}), \label{MomentEM}\\
    \partial_t(\bm{E}\cdot\bm{B})=&
    \frac{1}{\beta_\star^2}\bm{B}\cdot(\nabla\times\bm{B})
    -\bm{E}\cdot(\nabla\times\bm{E})-\frac{\bm{J}\cdot\bm{B}}{\sigma}.
    \label{EdotB}
\end{align}
To preserve the constraint $\bm{E}\cdot\bm{B}=0$, the current parallel to the magnetic field ($\bm{J}\cdot\bm{B}$) is determined such that the right hand side of \eqref{EdotB} becomes always zero.
\begin{align}
\frac{\bm{J}\cdot\bm{B}}{\sigma}=    \frac{1}{\beta_\star^2}\bm{B}\cdot(\nabla\times\bm{B})
-\bm{E}\cdot(\nabla\times\bm{E}).
\end{align}
On the other hand, the evolution equation \eqref{MomentEM} for $\bm{E}\times\bm{B}$ can be solved instead of \eqref{AM} because
the electric field has only the perpendicular component and can be reproduced by $\bm{E}=\bm{B}\times(\bm{E}\times\bm{B})/|\bm{B}|^2$. 
Moreover, we can easily eliminate $\epsilon_\sigma\tilde{n}\bm{E}+\bm{J}\times\bm{B}$ in \eqref{MomentP} and \eqref{MomentEM} by using Ohm's law \eqref{Ohm}. Then, the right hand sides no longer include $\bm{J}$. Therefore, the cold  RHMHD equations are regarded as a dynamical system of 10 fields $(n,\bm{M}_{\rm P},\bm{M}_{\rm EM}, \bm{B})$, where $\bm{M}_{\rm P}=n\gamma\bm{v}$ and $\bm{M}_{\rm EM}=\bm{E}\times\bm{B}$, satisfying two constraints $\bm{M}_{\rm EM}\cdot\bm{B}=0$ and $\nabla\cdot\bm{B}=0$. The right hand sides of \eqref{C}, \eqref{MomentP}, \eqref{FM} and \eqref{MomentEM} are explicitly written in terms of $(n,\bm{M}_{\rm P},\bm{M}_{\rm EM}, \bm{B})$, using 
\begin{align}
    \bm{v}=& \frac{\bm{M}_{\rm P}/n}{\sqrt{1+ \beta_\star^2(\bm{M}_{\rm P}/n)^2}  }, \label{v_Mp}\\
    \bm{E}=&\bm{B}\times\bm{M}_{\rm EM}/|\bm{B}|^2. 
\end{align}
In this way, the RHMHD equations are numerically solvable in the cold plasma limit.

We also note that the force-free state can be obtained in the high magnetization limit $\sigma\rightarrow\infty$, where only \eqref{FM} and \eqref{MomentEM} are solved by neglecting the last term of \eqref{MomentEM}.
Although $\bm{J}$ is already eliminated from these equations, it must exist at orders $\bm{J}\times\bm{B}=O(1)$ and $\bm{J}\cdot\bm{B}=O(\sigma)$. Our normalization $\bm{J}=O(1)$ is therefore violated in the parallel component and needs to be modified to treat this model.

On the other hand, in the case of RMHD which neglects the Hall effect $\epsilon_H=0$,  
Ohm's law \eqref{Ohm} becomes $\bm{E}+\bm{v}\times\bm{B}=0$ that does not include $\bm{J}$. Therefore, we are forced to eliminate $\bm{J}$ by combining \eqref{MomentP} and \eqref{MomentEM} as follows
\begin{align}
\partial_t(n \gamma\bm{v} +\sigma\bm{E}\times\bm{B})=&
      - \nabla\cdot(n \gamma \bm{v}\bm{v}) 
      +
      \sigma\nabla\cdot\left(\frac{\bm{B}\bm{B}}{\beta_\star^2}+\bm{E}\bm{E}\right)\nonumber\\
      &-\sigma\nabla\left(\frac{|\bm{B}|^2}{2\beta_\star^2}+\frac{|\bm{E}|^2}{2}\right),\label{Moment}
\end{align}
where $\bm{M}_{\rm tot}:=n\gamma \bm{v}+\sigma\bm{E}\times\bm{B}$ is the total momentum of plasma and electromagnetic field. According to Ohm's law, $\bm{E}$ is always replaced by $-\bm{v}\times\bm{B}$. Thus, the RMHD equations are a dynamical system of 7 fields $(n,\bm{M}_{\rm tot},\bm{B})$ with a constraint $\nabla\cdot\bm{B}=0$. However, to calculate the right hand sides of \eqref{C}, \eqref{FM} and \eqref{Moment}, we need to write $\bm{v}$ in terms of $(n,\bm{M}_{\rm tot},\bm{B})$. It is well known that this is not analytically feasible and requires the use of a root-finding algorithm (such as the Newton-Raphson method), which is one of the most computationally expensive part of RMHD simulation.

For the case of hot plasma, the momentum equation \eqref{MomentP} is rewritten as
\begin{eqnarray}
    \partial_t( h \gamma^2\bm{v} )= -\nabla p - \nabla\cdot(h \gamma^2 \bm{v}\bm{v}) 
+\epsilon_\sigma \tilde{n}\bm{E}
+ \bm{J}\times\bm{B}, \label{MomentP_hot}
\end{eqnarray}
including the enthalpy density $h$ and pressure $p$. The evolution equation for the total energy $E_{\rm tot}$ should be solved additionally when the barotropic condition is not satisfied, and the primitive variables $({N},\bm{v},p,\bm{B})$ should be reconstructed from the time-evolving variables. Since $M_p = h\gamma^2 v$ depends on $N=n/\gamma$ even in the barotropic case $h(N)$, the formula \eqref{v_Mp} no longer determines $\bm{v}$ explicitly. Unfortunately, a root-finding algorithm is again necessary for RHMHD.

Therefore, in the presence of the Hall effect and the cold plasma limit, we can avoid using root-finding algorithm and the time-marching algorithm becomes straightforward while
the number of field variables increases from 7 to 10. The cold RHMHD equations are possibly solved at a lower cost than the RMHD equations.

Finally, in the presence of the electron-inertia effect $\epsilon_I^2\ne0$, Ohm's law is regarded as the evolution equation for $\gamma\bm{J}$. The number of field variables is 13 under one constraint $\nabla\cdot\bm{B}=0$, which is essentially the same as the original two-fluid equations.
The numbers of field variables and constraints are summarized in Talbe.~\ref{table: variables}. As we have remarked before, it is more natural in XMHD (and IMHD) to solve $\bm{A}$ instead of $\bm{B}$ under the constraint $\nabla\cdot\bm{A}=0$. Since cold plasma is assumed in this work for simplicity, one more field variable (such as pressure or temperature) would be added when temperature is not negligible.

\begin{table}[h]
\begin{center}
\renewcommand\arraystretch{1.2}
\begin{tabular}{|c|l|l|}
\hline
 \textbf{Model} & \textbf{Field variables} & \textbf{Constraints} \\
\hline \hline
  MHD & 7 $(n,\bm{v},\bm{B})$ & 1 ($\nabla\cdot\bm{B}=0$)\\
  \hline
HMHD & 7 $(n,\bm{v},\bm{B})$ & 1 ($\nabla\cdot\bm{B}=0$)\\
   \hline
IMHD & 7 $(n,\bm{v},\bm{B})$& 1 ($\nabla\cdot\bm{B}=0$)\\
 \hline
XMHD & 7 $(n,\bm{v},\bm{B})$& 1 ($\nabla\cdot\bm{B}=0$)\\
 \hline
 RMHD &  7 $(n,\bm{v},\bm{B})$& 1 ($\nabla\cdot\bm{B}=0$)\\
   \hline
 RHMHD & 10 $(n,\bm{v},\bm{B},\bm{E})$ & 2 ($\nabla\cdot\bm{B}=0$, $\bm{E}\cdot\bm{B}=0$)\\
     \hline
 W-RHMHD & 10 $(n,\bm{v},\bm{B},\bm{E})$& 2 ($\nabla\cdot\bm{B}=0$, $\bm{E}\cdot\bm{B}=0$)\\
  \hline
 W-RIMHD & 13 $(n,\bm{v},\bm{J},\bm{B},\bm{E})$ & 1 ($\nabla\cdot\bm{B}=0$)\\
    \hline
 W-RXMHD & 13 $(n,\bm{v},\bm{J},\bm{B},\bm{E})$ & 1 ($\nabla\cdot\bm{B}=0$)\\
\hline
\end{tabular}
\caption{Number of field variables for cold plasma (H = Hall, I = Inertial, X = eXtended, R = Relativistic, W-R = Weakly-Relativistic) }
\label{table: variables}
\end{center}
\end{table}

\section{conclusion}

In this paper, we have investigated the applicability of various MHD models to special relativistic plasmas, using the method of dominant balance in the two-fluid equations. To simplify the formulation and consideration, we have assumed cold plasma (the limit of zero temperature and pressure) because electromagnetic force, not pressure, is a dominant force in the MHD balance. Although there is no problem in including nondominant pressure effect, the case of relativistic pressure should be investigated as a future topic which might also breaks down the MHD balance (or the charge neutrality approximation). Similarly, externally-applied electric field is assumed to be absent because it is rarely dominant.

Under these assumptions, the relativistic two-fluid equations are nondimensionalized by eight representative scales, resulting in seven nondimensional parameters. For the electromagnetic force to be a dominant term, the six balances (1 to 6) are imposed as constraints among these parameters. Since the balances 3 and 4 are inequalities, the number of the nondimensional parameters is reduced to three $(\epsilon_m,\sigma,\beta_\star)$ satisfying an inequality $\beta_\star^2/\sigma\le1$. The parameter $\epsilon_m=v_\star/(L_\star\omega_{c\star})$ is smaller than $1$ if we focus on the flow dynamics slower than the cyclotron frequency $\omega_{c\star}$. The RMHD equations are obtained in the limit $\epsilon_m\rightarrow0$. By taking the mass ratio as an additional parameter, this parameter $\epsilon_m$ appears only through either $\epsilon_H=\tilde{\mu}\epsilon_m$ or $\epsilon_I^2=\mu^2\epsilon_m^2$ in the two-fluid equations. We have shown that the approximation of proper charge neutrality can be justified by neglecting the order of $\epsilon_\sigma\epsilon_I^2$ where $\epsilon_\sigma=\min(\sigma,1)$. When $\sigma\ll1$, this approximation naturally corresponds to the quasi-neutrality condition of non-relativistic MHD. All the reduced models, or the generalized MHD models, are derived by neglecting $O(\epsilon_\sigma\epsilon_I^2)$ while allowing for the Hall effect $O(\epsilon_H)$, electron-inertia effect $O(\epsilon_I^2)$ and relativistic effects $O(\beta_\star^2)$ and $O(\sigma)$. A special care is therefore needed when both the electron-inertia and relativistic effects are taken into account simultaneously, because their multiplication $O(\epsilon_\sigma\epsilon_I^2)$ is not negligible unless both $\sigma$ and $\epsilon_I^2$ are much smaller than $1$. Only for the weakly relativistic case $\sigma\ll1$, we can use the W-RXMHD and W-RIMHD models, where proper charge neutrality is still valid. If $\epsilon_\sigma\epsilon_I^2\ll1$ is not fulfilled, the two-fluid equations should be solved without any approximation.

The case of $\beta_\star^2/\sigma\ll1$ is often uninteresting because it is almost vacuum (i.e., the kinetic energy is much smaller than the energy of externally-applied magnetic field). In this paper, we set aside the force-free state for simplicity, which allows for only the existence of finite parallel current even in the vacuum limit. On the other hand, the inequality $\beta_\star^2/\sigma\le1$ indicates that the maximum velocity scale should be $\beta_\star=\sqrt{\epsilon_\sigma}$ which is understood as relativistic Alfv\'en ordering ($v_\star\simeq v_{A\star}$). Interesting MHD phenomena are expected in this velocity scale. By focusing on this velocity scale, the number of the nondimensional parameters is further reduced to two $(\epsilon,\sigma)$ which are related to the scales of number density $n_\star$, magnetic field $B_\star$ and length $L_\star$. We have illustrated the applicable ranges of the various MHD models in terms of these scales. For a low density case or in a small scale, it is shown that the charge neutrality condition is violated at an intermediate strength of magnetic field around $\sigma\sim 1$. 

We have also summarized the number of field variables for each generalized MHD model. The RHMHD model is shown to be a dynamical system of 10 fields $(n,\bm{v},\bm{B},\bm{E})$ satisfying two constraints ($\nabla\cdot\bm{B}=0$ and $\bm{E}\cdot\bm{B}=0$). This number 10 is different from 7 of the other non-relativistic MHD models and 13 of the original two-fluid model. Moreover, the cold RHMHD equations describe the time marching of variables $(n,n\gamma\bm{v},\bm{E}\times\bm{B},\bm{B})$ and the primitive variables $(n,\bm{v},\bm{B},\bm{E})$ can be written explicitly by them. Since the cold RMHD equations requires the root-finding algorithm to calculate $(n,\bm{v},\bm{B})$ from $(n,n\gamma\bm{v}+\sigma\bm{E}\times\bm{B},\bm{B})$, the cold RHMHD model has an advantage in that the time-marching algorithm is simpler and less expensive than RMHD at the expense of increasing the field variables from 7 to 10. Naturally, RHMHD is a higher fidelity model than RMHD since it includes the Hall effect, which is known to be important for magnetic reconnection process in electron-ion plasma. The application of RHMHD is therefore expected to be beneficial both theoretically and numerically for analyzing relativistic plasma phenomena.

\begin{acknowledgments}
We thank Y. Kawazura and K. Toma for helpful discussion. This work was supported by JST SPRING, Grant Number JPMJSP2114, a Scholarship of Tohoku University, Division for Interdisciplinary Advanced  Research and Education, and IFS Graduate Student Overseas Presentation Award.
\end{acknowledgments}

\bibliography{POP24-AR-01681}

\end{document}